\documentclass[12pt]{article}
\usepackage{geometry}             
\usepackage{graphicx}
\usepackage{amssymb}
\usepackage{amsmath}
\usepackage{epstopdf}
\usepackage{subfigure}
\usepackage{cite}

\usepackage[hyperindex=true,
          pdfstartview=FitH,
          bookmarksnumbered=true,
          bookmarksopen=true,
          citecolor=blue,
          linkcolor=blue,
          colorlinks=true,
          pdfborder=001,
          unicode]{hyperref}

\begin{document}

\title{Hairy black holes of Lovelock-Born-Infeld-Scalar gravity}
\author{Kun Meng\\
School of Science, Tianjin Polytechnic University, \\
Tianjin 300387, China\\
emails: \href{mailto:mengkun@tjpu.edu.cn}{\it mengkun@tjpu.edu.cn}
 }
\date{}                             
\maketitle

\begin{abstract}
In this paper we construct new hairy black holes of Lovelock-Born-Infeld-scalar gravity in diverse dimensions. We study horizon structures of the black holes and find the black holes may possess three horizons, two horizons or one horizon. We give the thermodynamic quantities and check the first laws. We study the thermal phase transition behaviors of the black holes in $T$-$S$ plane,  and find for even-dimensional black holes, in some ranges of the parameters Van der Waals-like phase transitions occur, in some ranges of the parameters, reentrant phase transitions (RPTs) appear. For odd-dimensional black holes no such phase transitions are found.
\end{abstract}

\section{Introduction}
It's known that general relativity (GR) is non-renormalizable, however, it was found that higher derivative corrections to Einstein gravity can lead to a power-counting renormalizable theory\cite{Stelle1977}, therefore, researchers are motivated to study high derivative gravities. Another reason to study higher order gravities is that,  modifying gravity is an alternative way to explain the accelerating expansion of our universe without introducing dark energy.
Among the high derivative gravities, Lovelock gravity\cite{Lovelock}, which consists of dimensionally continued Euler characteristics, is the unique theory that degenerates to GR naturally in four dimensions. The field equations of Lovelock gravity contain no more than 2nd order derivatives of metric and it is a theory of gravity free of ghost at linear level. It was found that Gauss-Bonnet gravity, which is the second order Lovelock gravity, emerges in the low energy limit of string theory\cite{Wheeler,Schwarz}. Since Lovelock gravity contains a lot of Lovelock coefficients
 which makes it difficult to write the solution in an explicit form, Ba\~{n}ados, Teitelboim and Zanelli proposed a choice of the Lovelock coefficients, the theory obtained is called dimensionally continued gravity (DCG)\cite{BTZ93}. Neutral and charged black hole solutions of DCG have been found in\cite{BTZ93,Crisostomo00,Cai98}, thermodynamics of the black holes have been studied in \cite{BTZ93,Crisostomo00,Cai98,Cai06,Aiello04,MiskovicOlea}.

It is found that the linear Maxwell theory does not always work well to explain electromagnetic phenomena.  In 1936, Heisenberg and Euler proposed a nonlinear electromagnetic theory to describe the phenomena of quantum electrodynamics\cite{Heisenberg}. Also in 1930's, Born and Infeld proposed a non-linear electrodynamics to cancel the self-energy divergence of electron, the theory is now known as Born-Infeld (BI) electrodynamics\cite{BI}. The authors of \cite{Fradkin,Tseytlin} found that BI action could be reproduced in the framework of string theory. Nowadays, BI theory has also been used vastly to study dark energy, holographic superconductor and holographic entanglement entropy\cite{0307177,Jing1,Jing2}, etc. Therefore, it is interesting to adopt BI electrodynamic field as a matter source to construct black hole solutions. The first solution of Einstein-Born-Infeld theory was found by Hoffmann\cite{Hoffmann}, the solution is devoid of essential singularity at the origin. Subsequently, many black hole solutions of gravity coupled to BI electromagnetic field with or without a cosmological constant were found \cite{Oliveira,FernandoKrug,Dey,CaiBI,DehghaniHendi,Panah1508,Meng,Meng1712}. Thermodynamics of the BI black holes have been studied in\cite{Mo,Panah1510,Panah1608,Panah1708}.

It's well-known that, no-hair theorems forbid the existence of asymptotically flat black holes of Einstein gravity coupled to scalar field conformally\cite{nohair,nohair2}. When the cosmological constant vanishes, black holes with conformal scalar hair in 4 dimensions was found but the scalar field configuration was found to be divergent at the horizon\cite{Bekenstein}. For spacetime with dimensions higher than four, black holes of this type simply do not exist at all\cite{Xanthopoulos}. When the cosmological constant is non-vanishing, conformal hairy black holes in 3 and 4 dimensions have been found \cite{Martinez1,Martinez2}, with the configurations of the scalar field being regular outside or on the horizon. However, until recently black holes of this type have not been found for $d>4$ where no-go results were reported\cite{Martinez3}.
Recently, the authors of \cite{1112.4112,1401.4987} proposed a new form of gravity coupled to scalar field conformally. With the constituent
\begin{align}
S^{\alpha\beta}_{\mu\nu}=\phi^2R^{\alpha\beta}_{\mu\nu}-2\delta^{[\alpha}_{[\mu}\delta^{\beta]}_{\nu]}\nabla_\rho\phi\nabla^\rho\phi-4\phi\delta^{[\alpha}_{[\mu}\nabla_{\nu]}\nabla^{\beta]}\phi
+8\delta^{[\alpha}_{[\mu}\nabla_{\nu]}\phi\nabla^{\beta]}\phi,
\end{align}
which transforms homogeneously, under the conformal transformations $g_{\mu\nu}\rightarrow \Omega^2g_{\mu\nu}$ and $\phi\rightarrow\Omega^{-1}\phi$, as $S^{\alpha\beta}_{\mu\nu}\rightarrow \Omega^{-4}S^{\alpha\beta}_{\mu\nu}$, they construct the conformally invariant action
\begin{align}
I_s=\int \mathrm{d}^dx \sqrt{-g}\left[\sum_{p=0}^{n-1}\frac{b_p}{2^p}\phi^{d-4p}\delta^{\mu_1\cdots\mu_{2p}}_{\nu_1\cdots\nu_{2p}}S^{\nu_1\nu_2}_{\mu_1\mu_2}\cdots S^{\nu_{2p-1}\nu_{2p}}_{\mu_{2p-1}\mu_{2p}}\right]\label{Is}.
\end{align}
Where $\delta^{\mu_1\cdots\mu_{2p}}_{\nu_1\cdots\nu_{2p}}$ is the generalized Kronecker delta of order $2p$
\begin{align}
\delta^{\mu_1\cdots\mu_{2p}}_{\nu_1\cdots\nu_{2p}}=\left|\begin{array}{ccc}
                                                           \delta^{\mu_1}_{\nu_1} & \cdots & \delta^{\mu_1}_{\nu_{2p}} \\
                                                           \vdots&   & \vdots \\
                                                           \delta^{\mu_{2p}}_{\nu_1} &  \cdots & \delta^{\mu_{2p}}_{\nu_{2p}}
                                                         \end{array}
\right|.
\end{align}
The theory (\ref{Is}) is the most general one of gravity coupled to scalar field conformally with 2nd order field equations both for metric and scalar field. When $d=4$ or $p=1$, the theory (\ref{Is}) reduces to gravity coupled to scalar field conformally with potential $V(\phi)=(\lambda/4!)\phi^4$ and  non-minimal coupling term $(-1/12) R\phi^2$\cite{1401.4987}. Hairy black holes of the theory (\ref{Is}) have been constructed in \cite{1401.4987,1501.00184,1508.03780,Mann1612,1703.01633}, the configurations of scalar field are regular everywhere outside or on the horizon. These are the first example of black holes with conformal scalar hair in $d>4$. The calculations of \cite{1501.00184,1508.03780} indicate that, compared to black holes with no scalar hair the hairy black holes are the thermodynamical favored configurations, i.e., they are thermodynamical stable. In this paper, we would like to generalize the black holes constructed in our previous work \cite{Meng1712} to contain scalar hair, and study thermodynamics of the black holes obtained.

Thermodynamics of black holes always appeals to people. In the pioneering work \cite{HawkingPage}, Hawking and Page investigated phase transition between Schwarzchild-AdS black hole and AdS vacuum. Subsequently, thermal phase transitions of black holes in inverse temperature-horizon($1/T$-$r_{+}$) plane and temperature-entropy($T$-$S$) plane were studied in Refs.\cite{Chamblin9902,Chamblin9904,Mahapatra16,Zeng,Kuang}, it was found that the phase transition behaviors of the black holes resemble the ones of Van der Waals liquid-gas systems.  Later on, by identify cosmological constant as pressure of the system\cite{0904.2765}, people studied thermal phase transitions of black holes in extended phase space and found Van der Waals-like phase transitions too\cite{Mann1612,Zou1612,Mann1205,Zou1702,Mo}. Recently, in extended phase space the author of \cite{Mann1208} found coexistence of small/intermediate/large black holes and phase transitions between them. The phase transitions resemble the RPTs, which are observed in  multicomponent fluid systems, gels, ferroelectrics, liquid crystals, and binary gases\cite{Narayanan}. Because of the higher curvature terms in Lovelock gravity, the expression of pressure is highly nonlinear, which makes it very difficult to study phase transitions in extended phase space. On the other hand, it was argued that the thermal phase transitions of black holes in $T$-$S$ plane and in $P$-$V$ plane are dual to each other\cite{Spallucci,Kuang}. Thus in the following we will study thermal phase transitions of the black holes constructed below in $T$-$S$ plane and see whether Van der Waals-like phase transitions or especially RTPs occur.

The paper is organized as, in section \ref{section2} we construct hairy black holes of Lovelock-Born-Infeld-scalar gravity and study horizon structures of the black holes. In section \ref{section3}, we present the thermodynamic quantities and check first law, then we study thermal phase transitions of the black holes in $T$-$S$ plane. We summarize our calculations in the last section.

\section{Black hole solutions\label{section2}}

The action of Lovelock gravity coupled to BI electromagnetic field and conformally coupled to scalar field is given by:
\begin{align}
I=&\frac{1}{16\pi G}\int \mathrm{d}^dx \sqrt{-g}\bigg[\sum_{p=0}^{n-1}\frac{1}{2^p}\delta^{\mu_1\cdots\mu_{2p}}_{\nu_1\cdots\nu_{2p}}\big(a_p R^{\nu_1\nu_2}_{\mu_1\mu_2}\cdots R^{\nu_{2p-1}\nu_{2p}}_{\mu_{2p-1}\mu_{2p}}\nonumber\\
&+16\pi G\; b_p\phi^{d-4p}S^{\nu_1\nu_2}_{\mu_1\mu_2}\cdots S^{\nu_{2p-1}\nu_{2p}}_{\mu_{2p-1}\mu_{2p}}\big)+4\pi G L(F)\bigg]\label{action},
\end{align}
where $G$ is the gravitational constant, $L(F)$ is the Lagrangian density of Born-Infeld electromagnetic field
\begin{align}
L(F)=4\beta^2\left(1-\sqrt{1+\frac{F_{\mu\nu}F^{\mu\nu}}{2\beta^2}}\right),
\end{align}
The coefficients $a_p$ in (\ref{action}) are arbitrary constants, in the special case of DCG, the $a_p$'s are chosen as \cite{Kuang,BTZ93}
\begin{align}
a_p=\left(
                 \begin{array}{c}
                   n-1 \\
                   p \\
                 \end{array}
               \right)\frac{(d-2p-1)!}{(d-2)!l^{2(n-p-1)}}.
\end{align}
Note that, DCG  becomes Born-Infeld gravity in even dimensions and becomes Chern-Simons gravity in odd dimensions\cite{BTZ93,Crisostomo00}.

Taking variation w.r.t. the metric we obtains the equations of motion (EOM)
\begin{align}
\frac{1}{16\pi G}\sum_{p=0}^{n-1}\frac{a_p}{2^{p+1}}\delta^{\nu\lambda_1\cdots\lambda_{2p}}_{\mu\rho_1\cdots\rho_{2p}}R^{\rho_1\rho_2}_{\lambda_1\lambda_2}\cdots R^{\rho_{2p-1}\rho_{2p}}_{\lambda_{2p-1}\lambda_{2p}}={T^{(M)}}^{\nu}_\mu+{T^{(S)}}^{\nu}_\mu, \label{EOMG}
\end{align}
with the energy momentum tensor
\begin{align}
{T^{(M)}}^{\nu}_\mu=\frac{1}{8}\delta^\nu_\mu L(F)+\frac{1}{4}\frac{F_{\mu\lambda}F^{\nu\lambda}}{\sqrt{1+\frac{F_{\rho\sigma}F^{\rho\sigma}}{2\beta^2}}},\label{energymomentum}
\end{align}
and
\begin{align}
{T^{(S)}}^{\nu}_\mu=\sum_{p=0}^{n-1}\frac{b_p}{2^{p+1}}\phi^{d-4p}\delta^{\nu\lambda_1\cdots\lambda_{2p}}_{\mu\rho_1\cdots\rho_{2p}}S^{\rho_1\rho_2}_{\lambda_1\lambda_2}\cdots S^{\rho_{2p-1}\rho_{2p}}_{\lambda_{2p-1}\lambda_{2p}},\label{Senergymomentum}
\end{align}
The EOM of electromagnetic field is given by
\begin{align}
\partial_\mu\left(\frac{\sqrt{-g}F^{\mu\nu}}{\sqrt{1+\frac{F^{\rho\sigma}F_{\rho\sigma}}{2\beta^2}}}\right)=0.\label{EOMEM}
\end{align}
The EOM of scalar field reads
\begin{align}
\sum_{p=0}^{n-1}\frac{(d-2p)b_p}{2^p}\phi^{d-4p-1}\delta^{(p)}S^{(p)}=0,\label{EOMscalar}
\end{align}
here $\delta^{(p)}S^{(p)}\equiv \delta^{\lambda_1\cdots\lambda_{2p}}_{\rho_1\cdots\rho_{2p}}S^{\rho_1\rho_2}_{\lambda_1\lambda_2}\cdots S^{\rho_{2p-1}\rho_{2p}}_{\lambda_{2p-1}\lambda_{2p}}$. If the Eq.(\ref{EOMscalar}) is imposed, the trace of energy momentum tensor vanishes, this is consistent with conformal invariance of theory (\ref{Is}). We are interested in static black hole solutions, the metric ansatz is taken as
\begin{align}
ds^2=-f(r)dt^2+\frac{dr^2}{f(r)}+r^2d\Omega_{d-2}^2,\label{ansatz}
\end{align}
where $d\Omega_{d-2}^2$ is the line element of $(d-2)$-sphere.

Under the electrostatic potential assumption, all other components of the field strength $F_{\mu\nu}$ vanish except $F_{rt}$. Solving (\ref{EOMEM}) we have
\begin{align}
F_{rt}=\frac{Q}{\sqrt{r^{2d-4}+\frac{Q^2}{\beta^2}}}.\label{strength}
\end{align}
If the scalar configuration is taken as
\begin{align}
\phi=\frac{N}{r},\label{scalarform}
\end{align}
the EOM of scalar field (\ref{EOMscalar}) is satisfied provided
\begin{align}
\sum_{p=1}^{n-1}pb_p\frac{(d-1)!}{(d-2p-1)!}N^{2-2p}&=0,\nonumber\\
\sum_{p=0}^{n-1}b_p\frac{(d-1)!(d^2-d+4p^2)}{(d-2p-1)!}N^{-2p}&=0.\label{constraint}
\end{align}
Substituting (\ref{ansatz}), (\ref{strength}) and (\ref{scalarform})   into (\ref{EOMG}) we obtain the black hole solutions
\begin{align}
f(r)=1+\frac{r^2}{l^2}-r^2\left(\frac{16\pi GM}{\Sigma_{d-2}r^{d-1}}+\frac{32\pi G H}{r^d}+\frac{\delta_{d,2n-1}}{r^{d-1}}-\Xi\right)^{\frac{1}{n-1}},\label{solution}
\end{align}
with
\begin{align}
H=\sum_{p=0}^{n-1}b_p\frac{(d-2)!}{(d-2p-2)!}N^{d-2p},
\end{align}
and
\begin{align}
\Xi=\frac{16\pi G \beta^2}{d-1}\bigg(1-\frac{\sqrt{Q^2r^4+\beta^2r^{2d}}}{\beta r^d}
+\frac{(d-2)Q^2}{(d-3)\beta^2r^{2d-4}}\cdot\;_2F_1\left[\frac{1}{2},\frac{d-3}{2(d-2)},\frac{7-3d}{4-2d},\frac{-Q^2r^{4-2d}}{\beta^2}\right]\bigg).
\end{align}
Where   $\Sigma_{d-2}$ is the volume of $(d-2)$-sphere. The additive constant $\delta_{d,2n-1}$ in (\ref{solution}) is chosen so that the black hole horizon shrinks to a point when $M\rightarrow0$\cite{Crisostomo00,Kuang}.

It's easy to note that, in the limit $H\rightarrow0$, the black holes (\ref{solution}) degenerate to black holes of DCG coupled to BI electromagnetic field, which were found in our previous work\cite{Meng1712}. The black holes reduce to charged black holes of DCG \cite{BTZ93,Cai98,Kuang} in the limit $\beta\rightarrow\infty$ and $H\rightarrow0$, and reduce to neutral black hole of DCG  when $Q\rightarrow0$ and $H\rightarrow0$.

It should be pointed out that the solutions (\ref{solution}) exist only for $D>4$, since for $D=4$ the constraint equations (\ref{constraint}) require all $b_p$ vanish. Therefore, the solutions  are not generalizations of $D=4$ solution. It is guessed that the solutions of $D>4$ and $D=4$ may belong to different branches of a larger set of solutions\cite{1401.4987}.

\begin{figure}[h]
\begin{center}
\includegraphics[width=.30\textwidth]{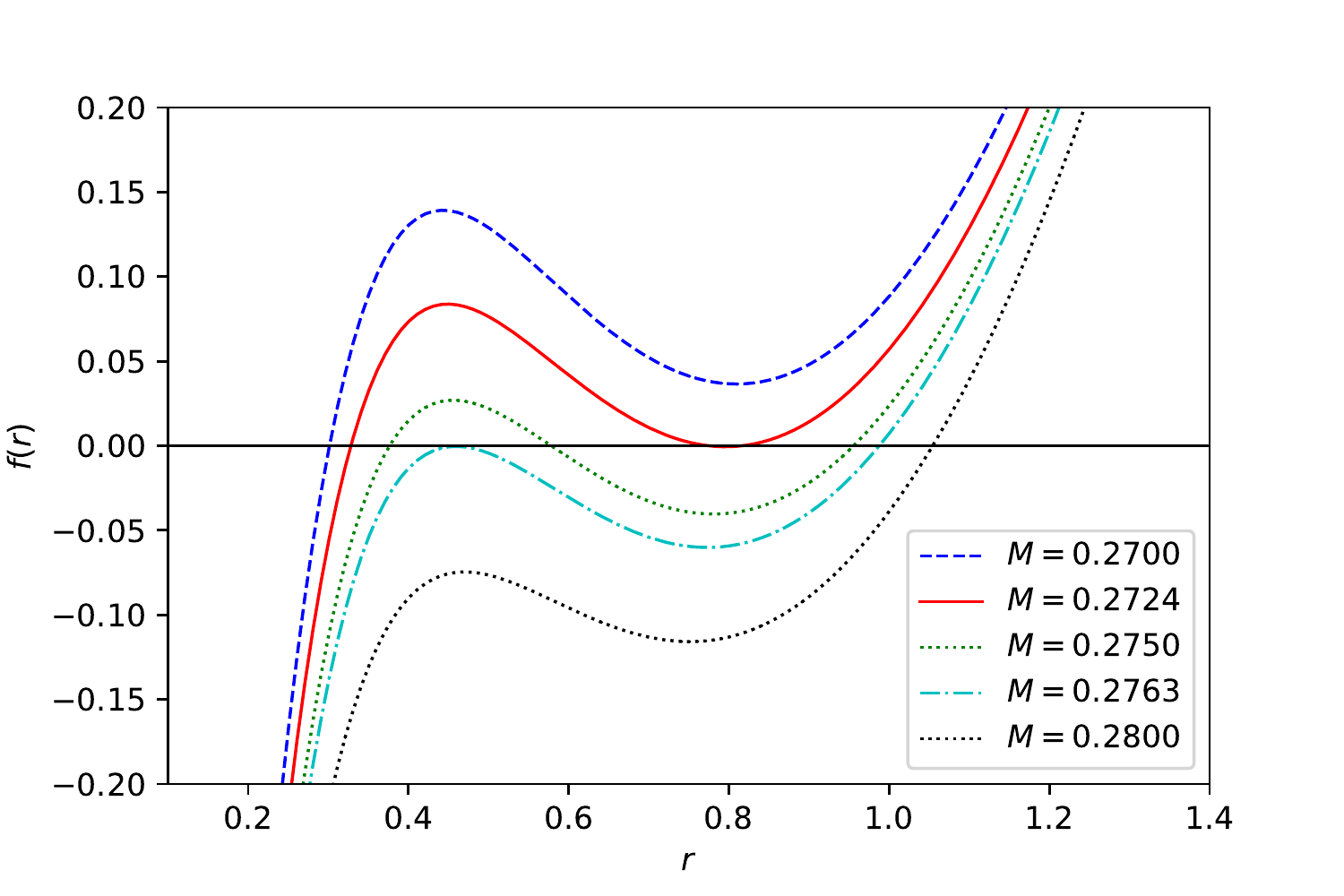}
\includegraphics[width=.30\textwidth]{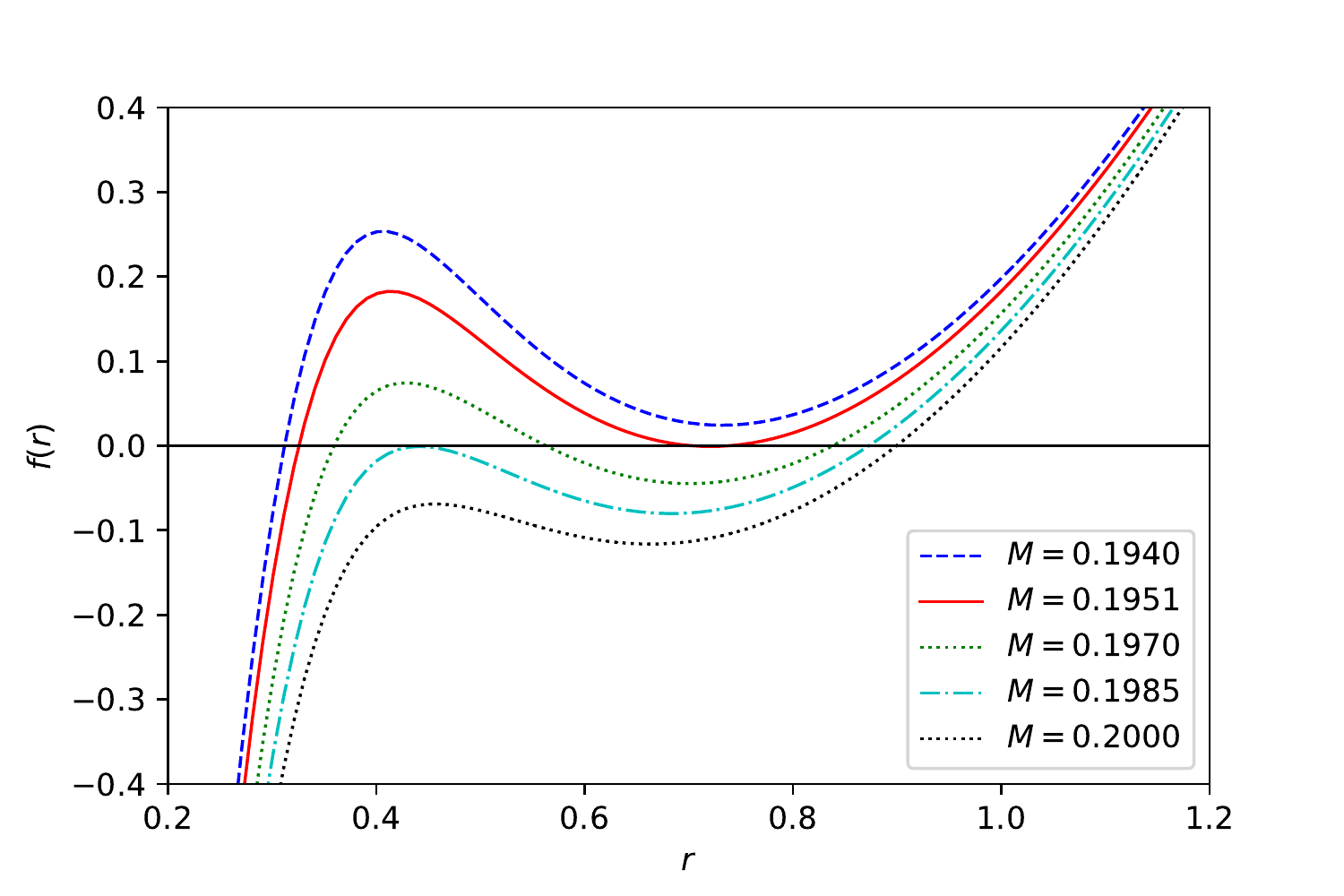}
\includegraphics[width=.30\textwidth]{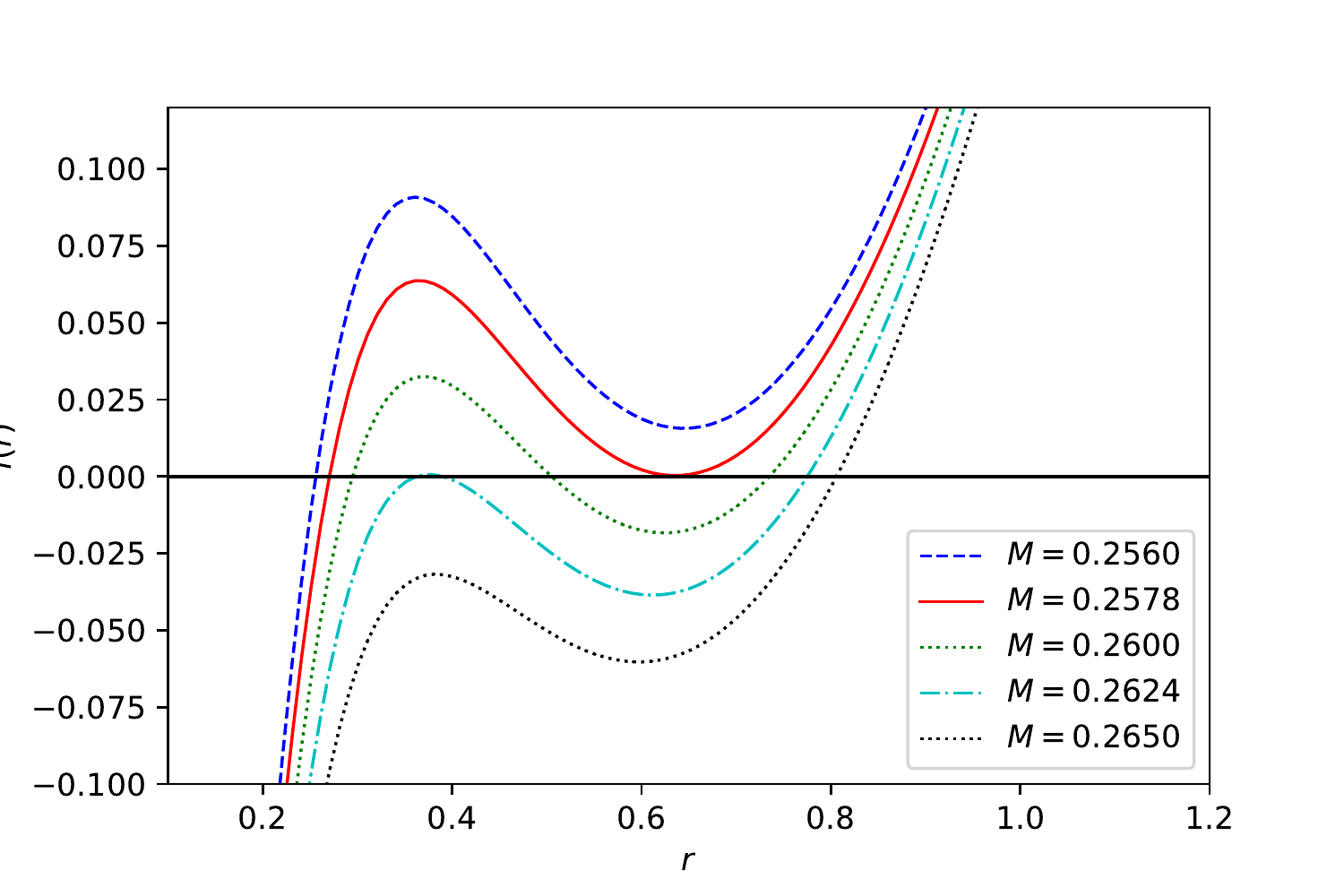}
\end{center}
\caption{$f(r)$ versus $r$. The parameters are fixed as $G=1,l=1,\beta=0.13$,  $Q=1.4$ (left), 1.5 (middle), 2.0 (right) and $d=5$ (left), 6 (middle), 7 (right).}
\label{fig1}
\end{figure}
Now, let's study horizon structures of the black holes. Fig.\ref{fig1} shows us the behaviors of $f(r)$ for fixed $Q$ with varying $M$. It's easy to see that, when $Q$ is held fixed and $M$ equals to some $M_{ext}$, it describes an extremal black hole, as the solid red line shows us. When $M$ decreases to below $M_{ext}$, it describes a Schwarzschild-like black hole with single horizon, as the dashed blue line shows. When $M$ increases to above $M_{ext}$, the black hole  may have three horizons, as displayed by the dotted green line. As $M$ go on increasing, the black holes may have two horizons, as the dash-dotted cyan line shows, or one horizon, as the dotted black line shows.

Owing to the scalar hair, the horizon structures of the black holes found here are different from the ones of the black holes we found before\cite{Meng1712}. Here the black holes may possess three horizons, in \cite{Meng1712} the black holes with more than two horizons were not found. $f(r)$ here is always divergent at $r=0$, while $f(r)$ in \cite{Meng1712}  may be finite at the origin for some values of the parameters.

\section{Thermodynamics\label{section3}}
In this section, we study  thermodynamics of the black holes. First we give the thermodynamic quantities and obtain the first law of thermodynamics, then we study thermal phase transitions of the black holes in $T$-$S$ plane.

\subsection{First law of thermodynamics}
Mass of the black hole above, in terms of the horizon radius, is given by
\begin{align}
M=&\frac{\Sigma_{d-2}}{16 \pi G  r_{+}^{d+1}(d-1)(d-3)(l^2+r_{+}^2)}\bigg\{(3-d)r_{+}^d\bigg[(1-d)l^2r_{+}^{d+2}\left(\frac{1}{l^2}+\frac{1}{r_{+}^2}\right)^n\nonumber\\
&+16\pi G (l^2+r_{+}^2)\left(2(d-1)H-\beta\left(r_{+}^d\beta-\sqrt{Q^2r_{+}^4+r_{+}^{2d}\beta^2}\right)\right)\bigg]\nonumber\\
&+16(d-2)\pi G Q^2r_{+}^4(l^2+r_{+}^2)\;_2F_1\left[\frac{1}{2},\frac{d-3}{2(d-2)},\frac{7-3d}{4-2d},\frac{-Q^2r_{+}^{4-2d}}{\beta^2}\right]\bigg\}-\frac{\delta_{d,2n-1}\Sigma_{d-2}}{16\pi G}
.\label{mass}
\end{align}
Temperature of the black hole, which is given by Hawking temperature $T=\frac{f^{'}(r_+)}{4\pi}$, reads
\begin{align}
T&=\frac{\left(\frac{1}{l^2}+\frac{1}{r_{+}^2}\right)^{-n}}{4l^4(n-1)\pi r_{+}^{d+3} \sqrt{Q^2r_{+}^4+r_{+}^{2d}\beta^2}}\bigg[l^2r_{+}^{d+2}\left(\frac{1}{l^2}+\frac{1}{r_{+}^2}\right)^n
\big(l^2(d+1-2n)\nonumber\\
&\;\;\;\;+(d-1)r_{+}^2\big) \sqrt{Q^2r_{+}^4+r_{+}^{2d}\beta^2}+32\pi G H (l^2+r_{+}^2)^2 \sqrt{Q^2r_{+}^4+r_{+}^{2d}\beta^2}\nonumber\\
&\;\;\;\;-16\pi G \beta (l^2+r_{+}^2)^2\left(Q^2r_{+}^4+r_{+}^{d}\beta\left(r_{+}^{d}\beta-\sqrt{Q^2r_{+}^4+r_{+}^{2d}\beta^2}\right)\right)\bigg].
\end{align}
Wald entropy of the black hole is
\begin{align}
S&=-2\pi\oint d^{d-2}x\sqrt{h}Y^{\mu\nu\rho\sigma}\epsilon_{\mu\nu}\epsilon_{\rho\sigma}\nonumber\\
&=\frac{\Sigma_{d-2}r_{+}^{d-2}}{2G}\sum_{p=1}^{n-1}\frac{p}{2^p}\delta^{\mu_1\cdots\mu_{2p-2}}_{\nu_1\cdots\nu_{2p-2}}\left(a_pR^{\nu_1\nu_2}_{\mu_1\mu_2}\cdots R^{\nu_{2p-4}\nu_{2p-2}}_{\mu_{2p-4}\mu_{2p-2}}+b_p\phi^{d-4p+2}S^{\nu_1\nu_2}_{\mu_1\mu_2}\cdots S^{\nu_{2p-4}\nu_{2p-2}}_{\mu_{2p-4}\mu_{2p-2}}\right)\nonumber\\
&=\frac{\Sigma_{d-2}}{4G}\sum_{p=1}^{n-1}pb_p\frac{(d-2)!}{(d-2p)!}N^{d-2p}\nonumber\\
&\;\;\;\;+\frac{(n-1)\Sigma_{d-2}r_+^d}{4 G(d-2n+2)}\left(\frac{1}{r_+^2}+\frac{1}{l^2}\right)^{n-1}\;_2F_1\left[1,\frac{d}{2},\frac{d-2n+4}{2},-\frac{r_+^2}{l^2}\right].
\end{align}
Where $\epsilon_{\mu\nu}$ is the normal bivector of  the $t=const$ and $r=r_{+}$ hypersurface with $\epsilon_{\mu\nu}\epsilon^{\mu\nu}=-2$, and $Y^{\mu\nu\rho\sigma}=\frac{\partial \mathcal{L}}{\partial R_{\mu\nu\rho\sigma}}$. The electric charge is calculated through performing integration of the flux of electromagnetic field in (\ref{EOMEM}) on the $t=const$ and $r\rightarrow\infty$ hypersurface, yielding
\begin{align}
Q_e=Q\Sigma_{d-2}.
\end{align}
The electric potential is given by
\begin{align}
\Phi&=A_\mu\chi^\mu|_{r\rightarrow\infty}-A_\mu\chi^\mu|_{r=r_{+}}\nonumber\\
&=\frac{Q}{(d-3)r_{+}^{d-3}}\cdot\;_2F_1\left[\frac{1}{2},\frac{d-3}{2(d-2)},\frac{7-3d}{2(2-d)},-\frac{Q^2r_{+}^{4-2d}}{\beta^2}\right]
\end{align}

With the thermodynamical quantities above,  the first law of thermodynamics
\begin{align}
  dM=TdS+\Phi dQ_e
\end{align}
can be check to be satisfied.

\begin{table}
\centering
\begin{tabular}{cccccc}
\hline\hline
$n$ &$\beta$ & $H$ &$S_{c}$ & $Q_c$ & $T_c$ \\ [0.5ex]
\hline
3& 9.80 &  0.01 & 0.146312 & 0.0951396  &  0.2382360  \\
\hline
3&10.0 &  0.01  & 0.146321  & 0.0951375  &  0.2382370  \\
\hline
3& 10.2 &0.01  & 0.146329 &  0.0951355 &  0.2382380  \\
\hline
4&9.80 & 0.01 & 0.240208 & 0.0470022 & 0.1946574  \\
\hline
4&10.0 &  0.01 & 0.240226&  0.0470001 & 0.1946575  \\
\hline
4&10.2 &  0.01 & 0.240242 &  0.0469982 &  0.1946580 \\
\hline
5&9.80 & 0.01 & 0.325397 &  0.0216006 &  0.1682470 \\
\hline
5& 10.0 & 0.01  &  0.325459 & 0.0215975  &  0.1682480 \\
\hline
5& 10.2&  0.01 & 0.325518 &0.0215946  & 0.1682490  \\
\hline
\end{tabular}
\label{table1}
\caption{Critical values of entropy, electric charge and temperature with $G=l=1$.}
\end{table}

\subsection{Thermal phase transition}
Now we study phase transitions of the black holes in the $T$-$S$ plane with $Q$  kept fixed.  The critical equations are given by
\begin{align}
\frac{\partial T}{\partial S}\bigg|_Q=0,\;\;\;\;\;\;\;\;\;\;\;\;\;\frac{\partial^2 T}{\partial S^2}\bigg|_Q=0.\label{criticaleq}
\end{align}
Note that the derivative $\frac{\partial T}{\partial S}\big|_Q$ is performed through $\frac{\frac{\partial T}{\partial r_{+}}}{\frac{\partial S}{\partial r_{+}}}\bigg|_Q$. First let's consider the even-dimensional black holes ($d=2n$). Solving the equation $\frac{\partial T}{\partial S}\big|_Q=0$ we have
\begin{align}
Q^2=-r_{+}^{4(n-1)}\beta^2+\frac{\left(\Upsilon+\sqrt{2048(n-1)\pi^2G^2\beta^4r_{+}^{4n}(l^2+r_{+}^2)^3
(l^2+(2n-3)r_{+}^2)+\Upsilon^2}\right)^2}{1024\pi^2G^2\beta^2r_{+}^4(l^2+r_{+}^2)^2(l^2+(2n-3)r_{+}^2)^2},\label{criticalQ2}
\end{align}
with
\begin{align}
\Upsilon=&l^2r_{+}^{2(n+1)}\left(\frac{1}{l^2}+\frac{1}{r_{+}^2}\right)^n(l^2+(1-2n)r_{+}^2)\nonumber\\
&+16\pi G(l^2+r_{+}^2)\left(6Hl^2+2H(2n-1)r_{+}^2-r_{+}^{2n}\beta^2(l^2(2n-3)+r_{+}^2)\right).
\end{align}
By inserting $Q^2$ in (\ref{criticalQ2}) into $\frac{\partial^2 T}{\partial S^2}\big|_Q=0$, after some variable substitutions,  fortunately we obtain the explicit form of the critical equation, which is a little lengthy we will not present it here. The critical equation is too complicated to be solved analytically, numeral results are listed in the tables.  From table 1 one can see that, when $H$ is fixed, as $\beta$ increases, the critical entropy and temperature increase while the critical electric charge decreases. From table 2 one sees that, when $\beta$ is fixed, as $H$ increases, the critical entropy, the critical temperature and the critical electric charge all increase. It's also noted from tables 1 and 2 that as the dimension of spacetime increases, the critical entropy increases while the critical temperature and critical electric charge decrease. For odd-dimensional black holes ($d=2n-1$) we repeat the above procedure, however, no phase transitions are found. This agrees with the results found in Refs.\cite{Kuang} and \cite{Meng1712}, there thermal phase transitions of odd-dimensional black holes in $T$-$S$ plane were not found too.

\begin{table}
\centering
\begin{tabular}{cccccc}
\hline\hline
$n$ &$\beta$ & $H$ &$S_{c}$ & $Q_c$ & $T_c$ \\ [0.5ex]
\hline
3& 10 &  0.008 &0.132692 & 0.0818376 &  0.231887  \\
\hline
3& 10 &  0.010  & 0.146321  & 0.0951375  &  0.238237  \\
\hline
3& 10 &0.012  & 0.158831 &  0.1076940 &  0.243736  \\
\hline
4&10  & 0.008 & 0.215908 & 0.0387992 & 0.189762   \\
\hline
4&10 &  0.010 & 0.240226 &  0.0470001 &0.194657  \\
\hline
4&10 &  0.012 & 0.262599 &  0.0550068 &  0.198840 \\
\hline
5&10 & 0.008  & 0.290659 &  0.0171849 & 0.164223 \\
\hline
5& 10& 0.010  &  0.325459 & 0.0215975 &  0.168248 \\
\hline
5& 10&  0.012 & 0.357570 & 0.0260368  & 0.171660 \\
\hline
\end{tabular}
\label{table2}
\caption{Critical values of entropy, electric charge and temperature with $G=l=1$.}
\end{table}

\begin{figure}[h]
\begin{center}
\includegraphics[width=.32\textwidth]{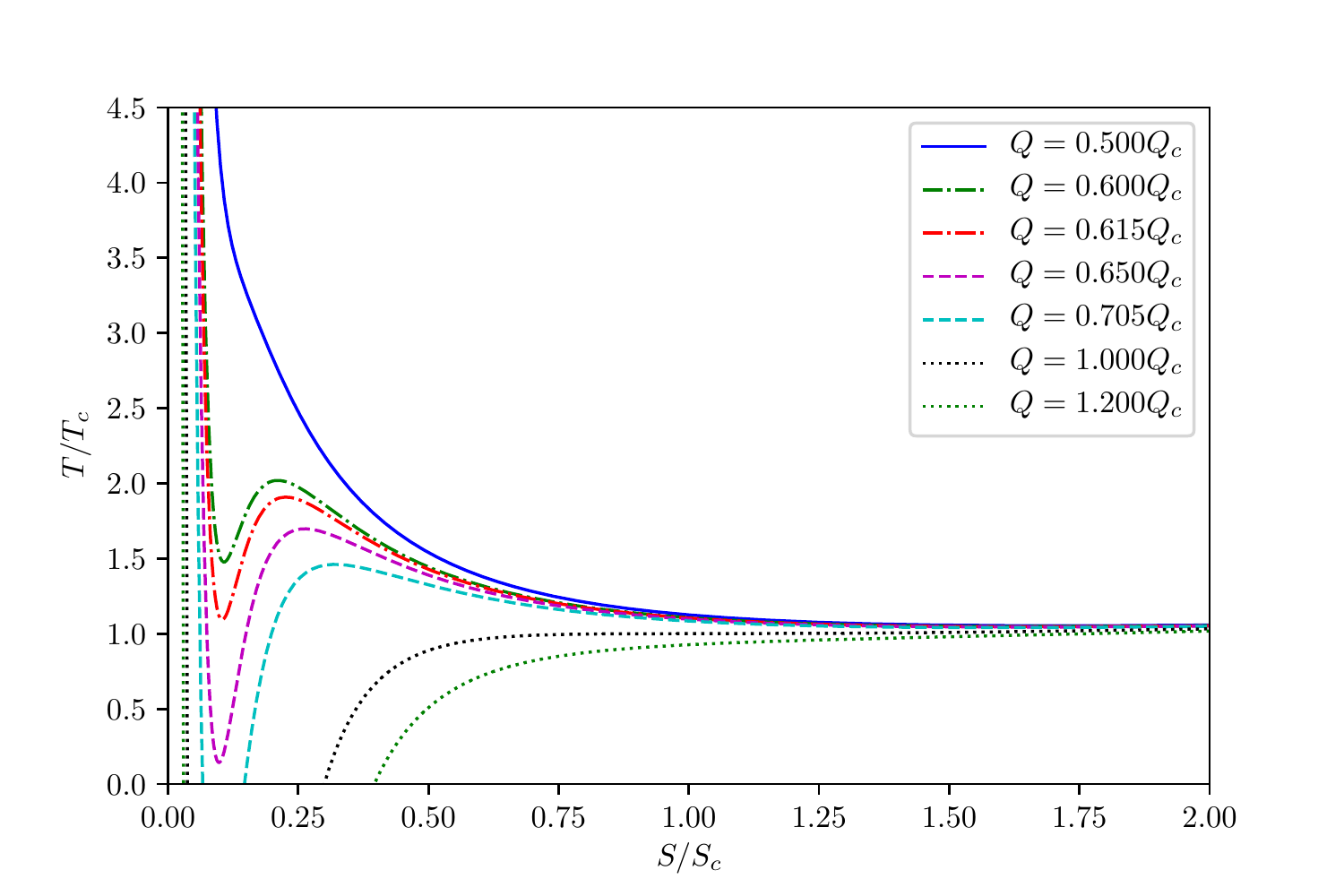}
\includegraphics[width=.32\textwidth]{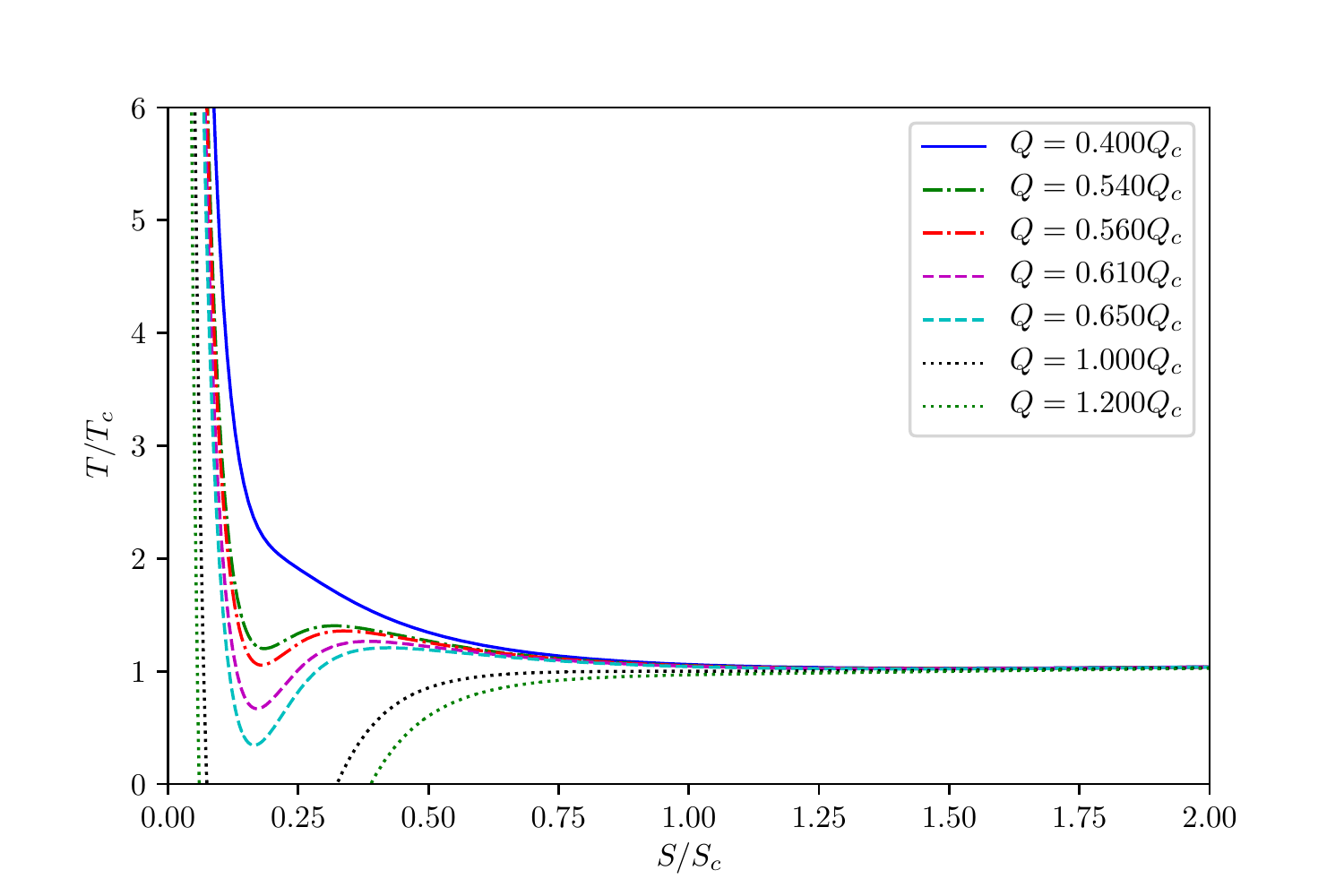}
\includegraphics[width=.32\textwidth]{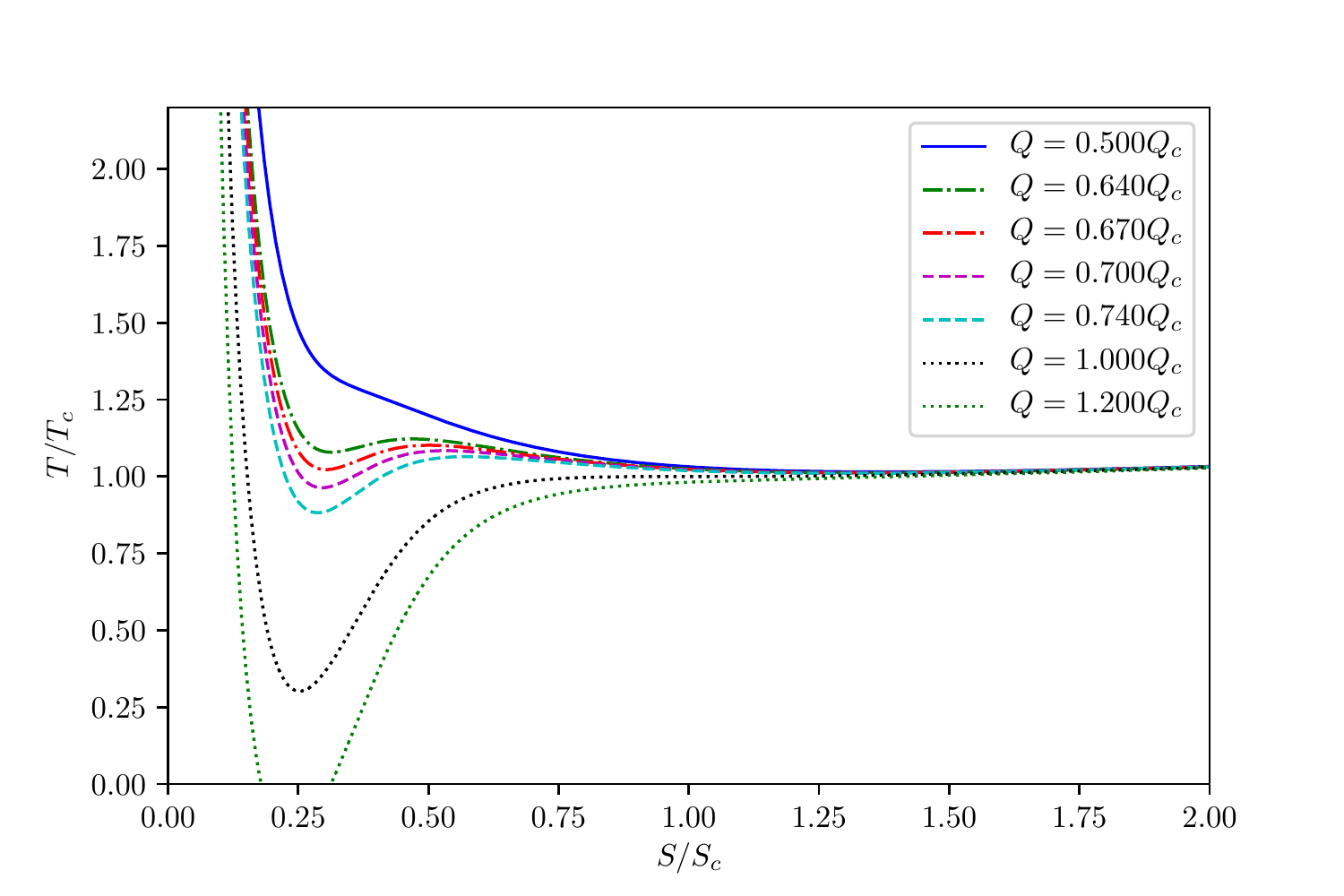}
\end{center}
\begin{center}
\includegraphics[width=.32\textwidth]{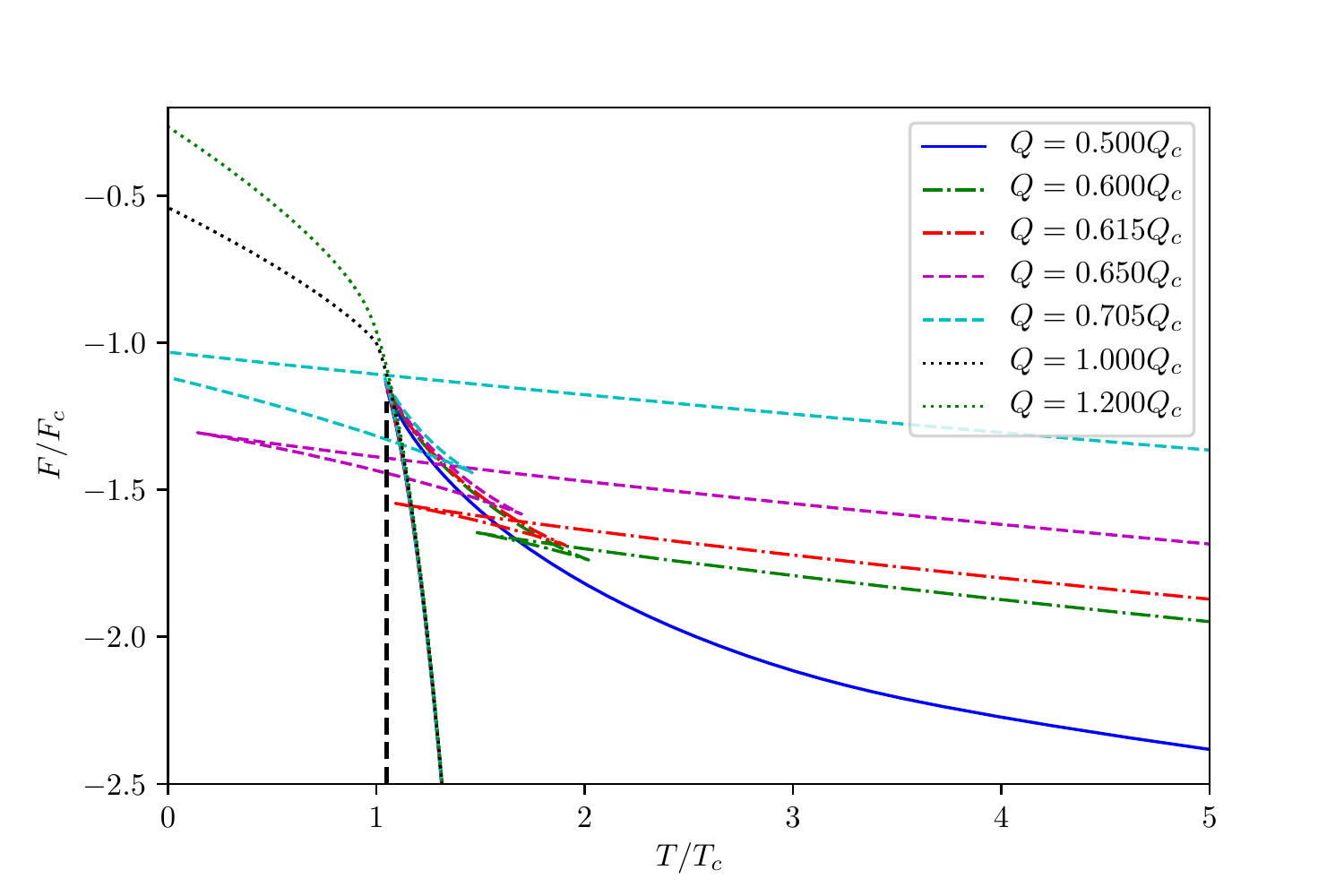}
\includegraphics[width=.32\textwidth]{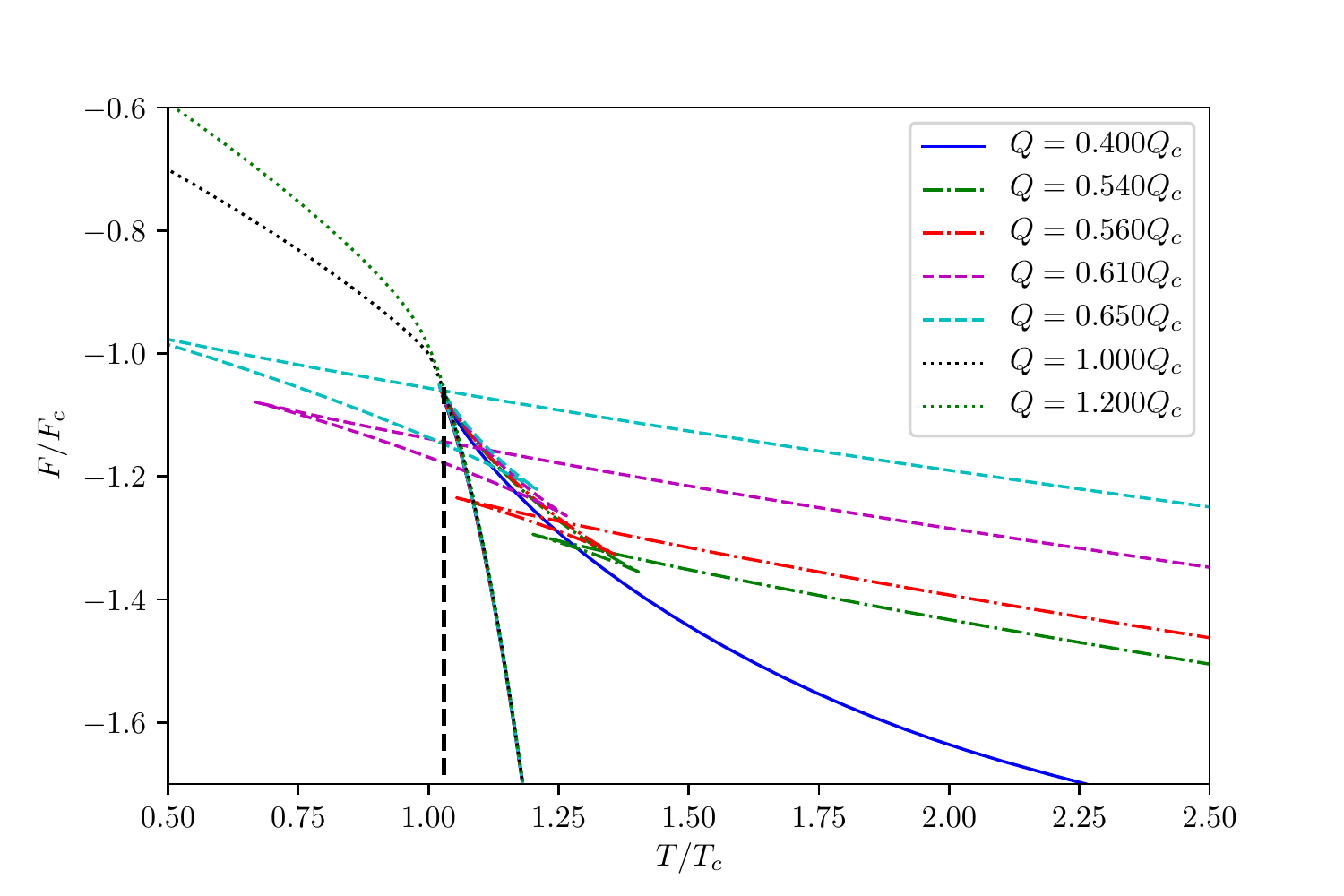}
\includegraphics[width=.32\textwidth]{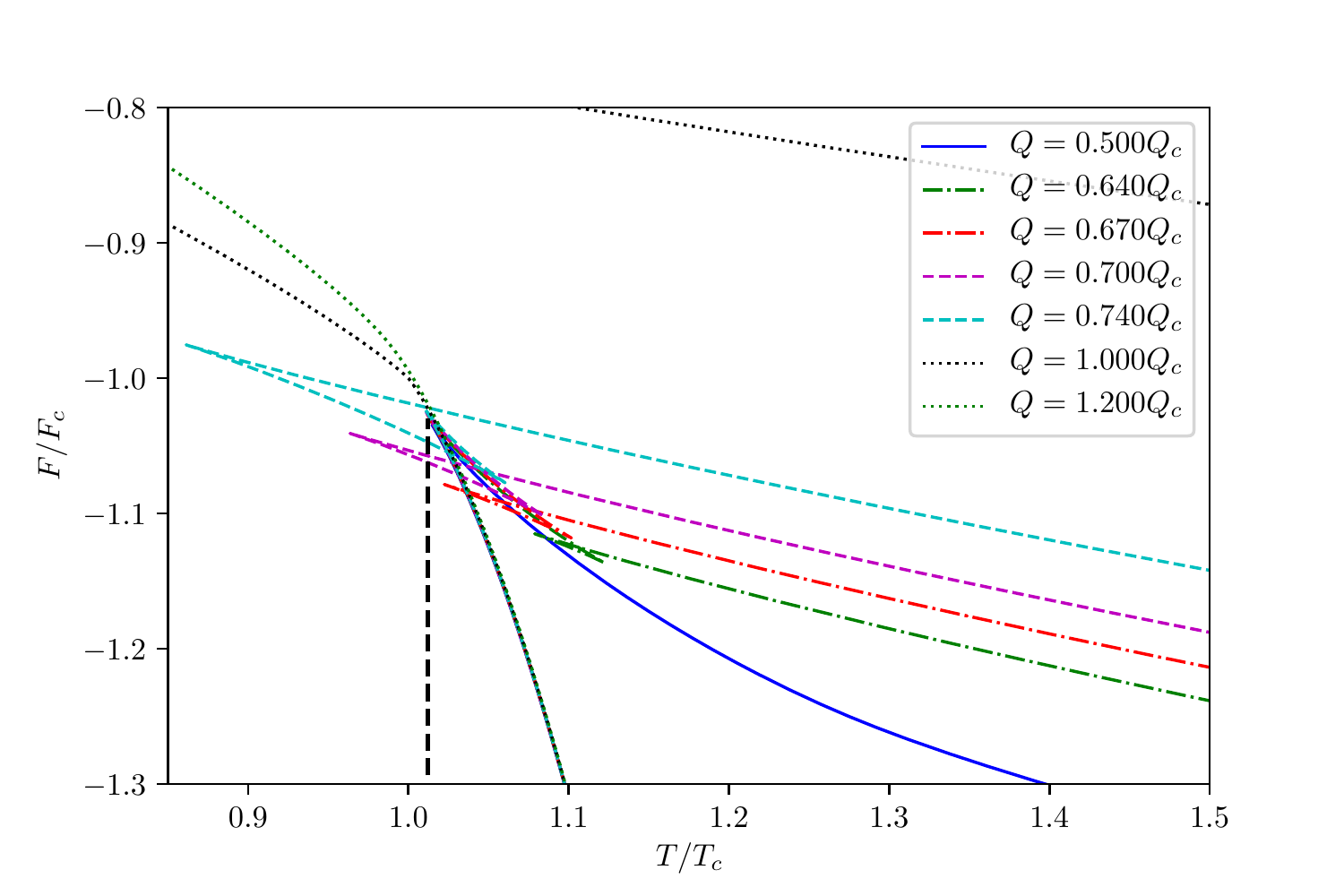}
\end{center}
\caption{Isocharge plots for black holes with $d=6$ (left two), $d=8$ (middle two) and $d=10$ (right two). The parameters are fixed as $G=1, H=0.01, \beta=10, l=1$.}
\label{fig2}
\end{figure}

\begin{figure}[h]
\begin{center}
\includegraphics[width=.5\textwidth]{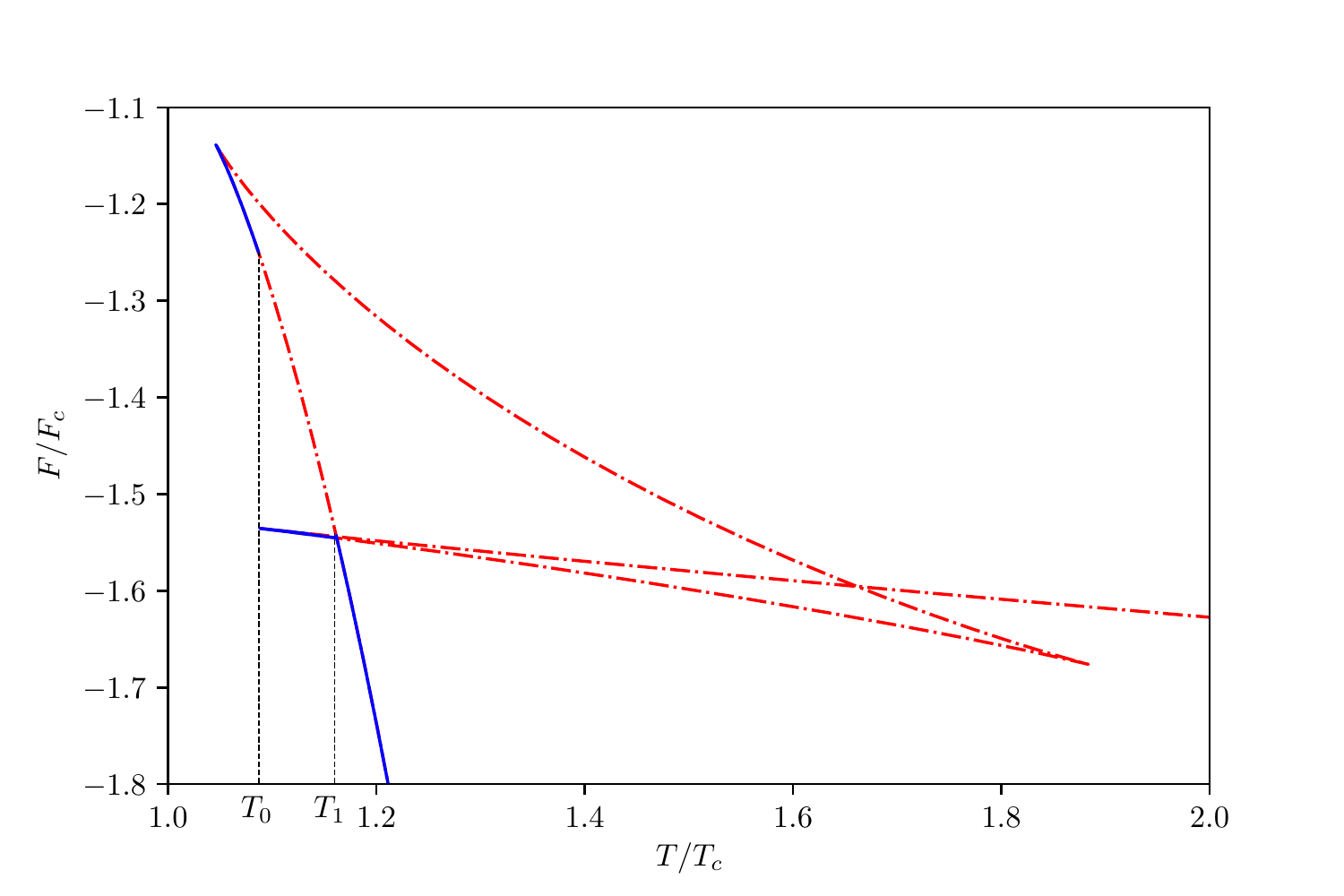}
\end{center}
\caption{A close-up of the dash-dotted red lines in $F$-$T$ planes in Fig.\ref{fig2}. At $T=T_0$ there exists an ``zero order phase transition''. At $T=T_1$ there exists a first order phase transition.}
\label{fig3}
\end{figure}
In Fig.\ref{fig2}, we present the isocharge plots of 6-dimensional($6d$), $8d$ and $10d$ black holes in $T$-$S$ and $F$-$T$ planes, where $F$ is the free energy which is defined as $F=M-TS$. From Fig.\ref{fig2} one can see that, for fixed $\beta$ and $H$, when $Q>Q_c$ the black hole is always stable, no phase transition appears, as the dotted black and green lines show us. When $Q$ decreases to below $Q_c$, large/small black holes  phase transition occurs, this phase transition is a first order one, as displayed by the cyan and purple lines. When $Q$ goes on decreasing, there exists a range of $Q$, $Q\in(Q_t, Q_z)$, for which the free energy is discontinuous at some $T=T_0$ (see Fig.\ref{fig3}), at which the black hole undergoes a ``zero order phase transition''  between  intermediate size and small size black holes. This phenomenon is also seen in superfluidity and superconductivity\cite{Maslov}. At some $T=T_1$ the black hole undergoes a  first order phase transition between large/small black holes. This kind of phase transition is reminiscent of the RPT, as  the blue lines show us. When $Q$ decreases to below $Q_t$, no Van der Waals-like phase transition appears.

\section{Conclusions}

In this paper, we construct new hairy black holes of Lovelock-Born-Infeld-scalar gravity. We study horizon structures of the black holes and find the black holes may possess three horizons, two horizons or one horizon. Owing to the scalar hair, horizon structures of the black holes constructed in this paper are different from the ones of the black holes found by us in\cite{Meng1712}, there the black holes with more than two horizons were not found.  Metric singularities at the origin always exist for black holes in this paper, while there may be no metric singularities at the origin for black holes in \cite{Meng1712} in some ranges of the parameters.

We study thermal phase transitions of the black holes in $T$-$S$ and $F$-$T$ planes with fixed $Q$, and find there exist thermal phase transitions for even-dimensional black holes while no phase transitions occur for odd-dimensional black holes. For even-dimensional black holes we find that, when $Q>Q_c$ the black holes are stable, no phase transitions appear. When $Q$ decreases to below $Q_c$, Van der Waals-like phase transitions occur. When $Q$ goes on decreasing, there exists a range of $Q$, $Q\in(Q_t, Q_z)$, in this range the black holes undergo a ``zero order phase transition'' at some $T=T_0$ and undergo a first order Van der Waals-like phase transition at some $T=T_1$, i.e., RPTs are found in $T$-$S$ plane. When $Q$ decreases to below $Q_t$, no Van der Waals-like phase transitions appear.

\section*{Acknowledgment}

KM would like to thank Professor Liu Zhao for useful discussions. This work is supported by the National Natural Science Foundation of China (NSFC) under the
grant numbers 11447153.

\providecommand{\href}[2]{#2}\begingroup
\footnotesize\itemsep=0pt
\providecommand{\eprint}[2][]{\href{http://arxiv.org/abs/#2}{arXiv:#2}}

\end{document}